\documentclass[12pt]{article}
\usepackage{epsfig,amsmath,amssymb,stmaryrd,enumerate}
\usepackage{graphicx}

\advance \textheight by \topskip
\advance \textheight by 1pt
\makeatother

\begin{document}

\title{
Periodic and rapid decay rank two self-adjoint commuting differential operators
}

\author{{Andrey E. Mironov}\thanks{This work was partially supported by
grant 12-01-33058 from the Russian Foundation for Basic Research; grant MD-5134.2012.1
from the President of Russia; and a grant from Dmitri Zimin's "Dynasty" foundation.}}
\date{}

\maketitle

\centerline{{\it
Dedicated to Sergey Petrovich Novikov on his 75th birthday}
}

\begin{quote}
\noindent{\sc Abstract. }
Self-adjoint rank two commuting ordinary differential operators are studied in this paper. Such operators with trigonometric, elliptic and
 rapid decay coefficients corresponding to hyperelliptic
spectral curves are constructed. Some problems related to the Lam\'e operator and rank two solutions of soliton
equations are discussed.
 \medskip
 \end{quote}

\section{Introduction}
Almost all solutions of soliton equations obtained in the last decades with the help of finite-gap theory
are rank one solutions. This means that eigenfunctions of auxiliary linear operators form linear bundles over spectral
curves. Meanwhile in  the outstanding papers of I.M. Krichever and S.P. Novikov \cite{KN1},\cite{KN2} rank $l>1$ solutions are investigated
(i.e. the eigenfunctions form vector bundles of rank $l$ over the spectral curves). One of the main difficulty to construct higher rank solutions (even
numerically) is the problem of finding higher rank commuting operators.

 In \cite{M4} an example of rank two operators corresponding to nonsingular spectral curves of arbitrary genus is constructed.
The operator
$$
 L^{^{\sharp}}_4=(\partial_x^2+\alpha_3x^3+\alpha_2 x^2+\alpha_1x+\alpha_0)^2+g(g+1)\alpha_3x, \qquad \alpha_3\ne 0
$$
commutes with a differential operator
$L_{4g+2}^{^{\sharp}}$ of order $4g+2$ \cite{M4}. The spectral curve is hyperelliptic curve of genus $g$.
In this paper we further develop  methods of \cite{M4} and construct self-adjoint operators of rank two with
trigonometric, elliptic and  rapid decay coefficients corresponding to hyperelliptic spectral curves.
 It is remarkable that there are higher rank operators with periodic and rapid decay coefficients expressed in terms of elementary functions
(see Corollaries 1 and 2 below).
The main results of this paper are Theorems 1 and 2.

\vspace{0.4cm}
\noindent{\bf Theorem 1} {\it The operator
$$
 L_4^{\natural}=(\partial_x^2+\alpha_1{\mathcal P}(x)+\alpha_0)^2+\alpha_1 g_2 g(g+1){\mathcal P}(x),\quad \alpha_1\ne 0,
$$
where ${\mathcal P}$ satisfies the equation
$$
 ({\mathcal P}'(x))^2=g_2{\mathcal P}^2(x)+g_1{\mathcal P}(x)+g_0,\quad g_2\ne 0
$$
commutes with an operator $L_{4g+2}^{\natural}$ of order $4g+2$.
}
\vspace{0.4cm}

\noindent At $g_0=1,g_1=0,g_2=-1$ we have

\vspace{0.4cm}
\noindent{\bf Corollary 1} {\it The operator with periodic coefficients
$$
 L_4^{\natural}=(\partial_x^2+\alpha_1\cos(x)+\alpha_0)^2-\alpha_1 g(g+1)\cos(x),\quad \alpha_1\ne 0
$$
commutes with an operator of order $4g+2$.
}
\vspace{0.4cm}

\noindent Let $\wp(z)$ be the Weierstrass elliptic function satisfying the equation
$$\
 (\wp'(z))^2=4 \wp^3(z)+g_2\wp^2(z)+g_1\wp(z)+g_0.
$$

\vspace{0.4cm}
\noindent{\bf Theorem 2} {\it The operator
$$
 L_4^{\flat}=(\partial_x^2+\alpha_1\wp(x)+\alpha_0)^2+s_1\wp(x)+s_2\wp^2(x),
$$
where
$$\alpha_1=\frac{1}{4}-2g^2-2g,\quad s_1=\frac{1}{4}g(g+1)(16\alpha_0+5g_2),\quad s_2=-4g(g+2)(g^2-1),$$
$\alpha_0$ is an arbitrary constant,
commutes with an operator of order $4g+2$.
}
\vspace{0.4cm}

\noindent At $g_2=4a^2$ we have

\vspace{0.4cm}

\noindent{\bf Corollary 2} {\it The operator with rapid decay coefficients
$$
 L_4^{\natural}=\left(\partial_x^2-\frac{\alpha_1a^2}{\cosh^2(ax)}+\alpha_0\right)^2-\frac{s_1a^2}{\cosh^2(ax)}+\frac{s_2a^4}{\cosh^4(ax)},\quad a\ne 0
$$
where
$$\alpha_1=\frac{1}{4}-2g^2-2g,\quad s_1=\frac{1}{4}g(g+1)(16\alpha_0+20a^2),\quad s_2=-4g(g+2)(g^2-1),$$
$\alpha_0$ is an arbitrary constant,
commutes with an operator of order $4g+2$.
}
\vspace{0.4cm}

\noindent Equations of spectral curves for pairs $L_4^{\flat},L_{4g+2}^{\flat}$ and $L_4^{\natural},L_{4g+2}^{\natural}$ are given in the Lemmas
1 and 2.

In the section 2 we recall the method of deformation of the Tyurin parameters.
In the section 3 we prove Theorems 1 and 2. In the Appendix I the spectral curve and eigenfunctions of the Lam\'e operator
$$
 L_2=-\partial_x^2+g(g+1)\wp(x)
$$
are found in some special form.
In the Appendix II rank two solutions of soliton equations are discussed.

The author is grateful to A.P. Veselov for valuable discussions.

\section{Krichever--Novikov theory (method of deformation of the Tyurin parameters)}

In this section we discuss some results on higher rank commuting ordinary differential operators and recall the method of deformation
of Tyurin parameters (applications of this method for finding higher rank solutions of soliton equations
see in \cite{KN1},\cite{KN2},\cite{Mokh}).

Commutative rings of ordinary differential operators were classified by I.M. Krichever \cite{K}.
Common eigenvalues of commuting operators
$$
 L_n=\partial_x^n+\sum_{i=0}^{n-2}u_{i}(x)\partial_x^i,\ \quad
 L_m=\partial_x^m+\sum_{i=0}^{m-2}v_{i}(x)\partial_x^{i}\
$$
are parametriezed by points of the spectral curve $\Gamma$. If
\begin{equation}\label{u}
 L_n\psi=z\psi,\qquad L_m\psi=w\psi,
\end{equation}
then $(z,w)\in\Gamma$, where $\Gamma$ is the smooth compactification of the curve defined by the equation $R(z,w)=0$, $R$ is the Burchnall--Chaundy \cite{BC}
polynomial such that $R(L_n,L_m)=0.$ {\it The rank of the pair $(L_n,L_m)$} is called the dimension of the space of common eigenfunctions (\ref{u}) for fixed
$P=(z,w)\in\Gamma$ in general position.
Eigenfunctions (Baker--Akhiezer function) and coefficients of rank one operators can be expressed via theta-function of $\Gamma$ \cite{K1}.
Operators of rank two corresponding to the elliptic spectral curves were found by I.M. Krichever and S.P. Novikov \cite{KN1},
operators of rank three with the same spectral curve were found by O.I. Mokhov \cite{Mokh}.
These operators and some related problems were studied in the papers \cite{GN}--\cite{Deh}.
Some examples of operators of rank 2 and 3 with spectral curves of genus 2,3,4 were found in \cite{M1}--\cite{Z}.

The case of rank 1 the Baker--Akhiezer function $\psi(x,P)$ has zeros divisor $\gamma(x)=\gamma_1(x)+\dots+\gamma_g(x)$ where $g$ is
genus of $\Gamma$. The $x$-trajectory of $\gamma$ in the Jacobi variety of $\Gamma$ is a straight line.
In the case of higher rank corresponding $x$-dynamics in the moduli space of vector bundles over $\Gamma$ is very complicated. Let me recall
spectral properties of $\psi=(\psi_1,\dots,\psi_l)$ at $l>1$ \cite{K}. The Baker--Akhiezer function $\psi$ has $lg$ simple poles
$p_1,\dots,p_{lg}$ with the properties
$$
 {\rm Res}_{p_i}\psi_j=v_{i,j}{\rm Res}_{p_i}\psi_{l},\quad 1\leq i\leq lg,\ 1\leq j\leq l-1.
$$
The set
\begin{equation}\label{pt}
 \{(p_j,v_{ik})\}
\end{equation}
 is called {\it Tyurin parameters}, which define a holomorphic bundle of rank $l$ over $\Gamma$.
There is a point $q\in\Gamma$ where $\psi$ has essential singularity of the form
$$
 \psi=\left(\sum_{s=0}^{\infty}\xi_s(x)k^{-s}\right)\Psi(x,k),
$$
where $\xi_0=(1,0,\dots,0), \xi_i(x)=(\xi_i^1(x),\dots,\xi_i^l(x))$, $k^{-1}$ is a local parameter,
the matrix  $\Psi$ satisfies the equation
$$
 \frac{d\Psi}{dx}=A\Psi,\quad
A=\left(
\begin{array}{cccccc}
 0 & 1 & 0 & \dots  & 0 & 0\\
 0 & 0 & 1 & \dots  & 0 & 0\\
 \dots & \dots & \dots & \dots &  \dots & \dots\\
 0 & 0 & 0 & \dots  & 0 & 1 \\
 k+\omega_0(x) & \omega_1(x) & \omega_2(x) & \dots & \omega_{l-2}(x) & 0
 \end{array}\right).
$$

The function $\psi$ can not be found explicitly. To find operators one can use the following
{\it method of deformation of Tyurin parameters}.
Let us consider for simplicity the case of rank two. Then for $P\in\Gamma$ the space of common eigenfunctions satisfies the second order
differential equation
$$
 \psi''-\chi_1(x,P)\psi'-\chi_0(x,P)\psi=0.
$$
The operator $\partial_x^2-\chi_1\partial_x-\chi_0$ is a common right divisor of
$L_n-z$ and $L_m-w$. Functions $\chi_0,\chi_1$ are rational functions on $\Gamma$, with poles at $p_1(x),\dots,p_{2g}(x)$ and $\chi_0$ has additional
pole at $q$, $\chi_1$ has zero at $q$
$$
 \chi_0=k+O(1),\quad \chi_1=O(1/k).
$$
Let $k-\gamma_i(x)$ be a local parameter near $p_i(x)$, then
$$
\chi_0(x,P)=\frac{-v_{i,0}(x)\gamma'_i(x)}{k-\gamma_i(x)}+d_{i,0}(x)+O(k-\gamma_i(x)),
$$
$$
\chi_1(x,P)=\frac{-\gamma'_i(x)}{k-\gamma_i(x)}+d_{i,1}(x)+O(k-\gamma_i(x)),
$$
where $d_{i,0},d_{i,1},\gamma_i(x),v_{i,0}(x)$ satisfy the Krichever equations \cite{K}
\begin{equation}\label{kr}
 d_{i,0}(x)=v_{i,0}^2(x)+v_{i,0}(x)d_{i,1}(x)-v'_{i,0}(x),\ i=1,\dots,2g.
\end{equation}
The last equations define deformations of Tyurin parameters and $(p_i(x),v_{i,0}(x))$ coincides with (\ref{pt}) at some $x=x_0$.
I.M. Krichever and S.P. Novikov found $\chi_0,\chi_1$ and corresponding operators at $g=1,l=2$.

\vspace{0.4cm}

\noindent{\bf Theorem 3} {\it (I.M. Krichever, S.P. Novikov \cite{KN1})
The operator
$$
 L_{KN}=\left(\partial_x^2+u\right)^2+2c_x(\wp(\gamma_2)-\wp(\gamma_1))\partial_x+(c_x(\wp(\gamma_2)-\wp(\gamma_1)))_x-
 \wp(\gamma_2)-\wp(\gamma_1),
$$
$$
 \gamma_1(x)=\gamma_0+c(x),\ \gamma_2(x)=\gamma_0-c(x),
$$
$$
 u(x)=-\frac{1}{4c_x^2}+\frac{1}{2}\frac{c_{xx}^2}{c_x^2}+2\Phi(\gamma_1,\gamma_2)c_x-\frac{c_{xxx}}{2c_x}+
 c_x^2(\Phi_c(\gamma_0+c,\gamma_0-c)-\Phi^2(\gamma_1,\gamma_2)),
$$
$$
 \Phi(\gamma_1,\gamma_2)=\zeta(\gamma_2-\gamma_1)+\zeta(\gamma_1)-\zeta(\gamma_2),
$$
where $c(x)$ is an arbitrary smooth function, $\gamma_0$ is a constant, commutes with an operator of order six.
}
\vspace{0.4cm}

The very interesting examples of the Krichever--Novikov operators are Dixmier operators.

\vspace{0.4cm}

\noindent{\bf Theorem 4} {\it (J. Dixmier  \cite{Dix}) The operators
$$
 L_{D}=\left(\frac{d^2}{dx^2}+x^3+h\right)^2+2x,\
$$
$$
 \tilde{L}_{D}=\left(\frac{d^2}{dx^2}+x^3+h\right)^3+\frac{3}{2}\left(x\left(\frac{d^2}{dx^2}+x^3+h\right)+
 \left(\frac{d^2}{dx^2}+x^3+h\right)x\right)
$$
commute, herewith
$$
 L_{D}^3=\tilde{L}_{D}^2-h.
$$
}

\noindent The action of the automorphisms group of the first Weyl algebra on the Dixmier operators
and on the operators $L_4^{\natural},L_{4g+2}^{\natural}$ were studied in \cite{Mokh3}.
 The conditions when $L_{KN}$ has rational coefficients were found in \cite{G}.

\vspace{0.4cm}

\noindent{\bf Theorem 5} {\it  (P.G. Grinevich \cite{G})
The operator $L_{KN}$ has  rational coefficients if and only if
$$
 c(x)=\int_{q(x)}^{\infty}\frac{dt}{\sqrt{P(t)}},
$$
where $q(x)$ is a rational function. If $\gamma_0=0$ and $q(x)=x$, then the operator $L_{KN}$ coincides with the Dixmier operator $L_D$.
}

\vspace{0.4cm}

\noindent The conditions when $L_{KN}$ is self-adjoint were found in \cite{GN}.

\vspace{0.4cm}

\noindent{\bf Theorem 6} {\it (P.G. Grinevich, S.P. Novikov \cite{GN})
Operator $L_{KN}$ is formally self-adjoint if and only if  $\wp(\gamma_1)=\wp(\gamma_2)$.
}

\vspace{0.4cm}

\noindent In \cite{M4} the operators of rank two corresponding to the hyperelliptic spectral curve
$$
 \Gamma: w^2=F_g(z)=z^{2g+1}+c_{2g}z^{2g}+\dots+c_0
$$
were considered
$$
 L_4\psi=z\psi,\quad L_{4g+2}\psi=w\psi, \quad  P=(z,w)\in\Gamma.
$$
\vspace{0.4cm}

\noindent{\bf Theorem 7} {\it (M. \cite{M4}) The operator $L_4$ is self-adjoint if
and only if
\begin{equation}\label{u2}
 \chi_1(x,P)=\chi_1(x,\sigma(P)),
\end{equation}
where $\sigma$ is the involution
$$
 \sigma(z,w)=(z,-w).
$$
}
\vspace{0.4cm}
Let us assume that $\chi_1$ is invariant under $\sigma$, then $L_4$ can be represented as
$$
 L_4=(\partial_x^2+V(x))^2+W(x)
$$
and $\chi_0,\chi_1$ have the form

\vspace{0.4cm}
\noindent{\bf Theorem 8}  {\it (M. \cite{M4}) Functions $\chi_0,\chi_1$ have the form
$$
 \chi_0=-\frac{1}{2}\frac{Q''}{Q}+\frac{w}{Q}-V, \qquad \chi_1=\frac{Q'}{Q},
$$
where $Q$ is a polynomial in $z$ of degree $g$ with coefficients depending on $x$
\begin{equation}\label{Q}
 Q(x,z)=(z-\gamma_1(x))\dots(z-\gamma_g(x)).
\end{equation}
 The polynomial $Q$ satisfies the equation
\begin{equation}\label{eq1}
 4F_g=4(z-W)Q^2-4V(Q')^2+(Q'')^2-2Q'Q^{(3)}
 +
 2Q(2V'Q'+4VQ''+Q^{(4)}),
\end{equation}
where $Q',Q'',Q^{(k)}$ mean $\partial_xQ,\partial_x^2Q,\partial_x^kQ.$}

\vspace{0.4cm}

\noindent{\bf Corollary 3}  {\it The function $Q$
satisfies the linear equation
\begin{equation}\label{r1}
 \mathcal{L}_5Q=\left(\partial_x^5+2\langle V,\partial_x^3\rangle+2\langle z-W-V'',\partial_x\rangle\right)Q=0,
\end{equation}
where $\langle A,B\rangle=AB+BA,$ the operator $\mathcal{L}_5$ is skew-symmetric. Potentials $V,W$ have the form
\begin{equation}\label{r2}
 V=\left(\frac{(Q'')^2-2Q'Q^{(3)}-4F_g(z)}{4(Q')^2}\right)\mid_{z=\gamma_j},
\end{equation}
$$
 W=-2(\gamma_1+\dots+\gamma_g)-c_{2g}.
$$
The functions $\gamma_1(x),\dots,\gamma_g(x)$ satisfy the equations
\begin{equation}\label{r3}
 \frac{(Q'')^2-2Q'Q^{(3)}-4F_g(z)}{4(Q')^2}\mid_{z=\gamma_j}=\frac{(Q'')^2-2Q'Q^{(3)}-4F_g(z)}{4(Q')^2}\mid_{z=\gamma_k}.
\end{equation}}
\vspace{0.4cm}

So, Theorem 8 and Corollary 3 show that the spectral theory of the operator $L_4$ is similar to the spectral theory of the finite-gap
one dimensional Schr\"odinger operator (see \cite{DMN})
$$
 (-\partial_x^2+u(x))\psi=z\psi.
$$
To the Schr\"odinger operator also corresponds a skew-symmetric operator but of order three \cite{DMN} (see also \cite{GD})
$$
 \mathcal{L}_3Q=\left(\partial_x^3+2\langle z-u,\partial_x\rangle\right)Q=0,
$$
and $Q$ satisfies the nonlinear equation
\begin{equation}\label{r4}
 4F_g(z)=4(z-u)Q^2-(Q')^2+QQ'',
\end{equation}
The potential $u$ has the form
$$
 u=-2(\gamma_1+\dots+\gamma_g)-c_{2g}.
$$
Equations (\ref{r3}) are analogues of the  Dubrovin equations
$$
 \gamma_j'=\pm\frac{2i\sqrt{F_g(\gamma_j)}}{\prod_{k\ne j}(\gamma_k-\gamma_j)}.
$$
Using the equation (\ref{eq1}) V.N. Davletshina \cite{Dav} proved that the operator
$$
 (\partial_x^2+\alpha_1e^x+\alpha_0)^2+g(g+1)\alpha_1e^x,\quad \alpha_1\ne 0
$$
commutes with an operator of order $4g+2$.

\section{Proof of Theorems}
\subsection{Theorem 2}
We construct the polynomial  $Q^{\natural}$ satisfying the equation (\ref{r1}) for
$$
 V^{\natural}=\alpha_1{\mathcal P}(x)+\alpha_0,\quad W^{\natural}=\alpha_1g_2 g(g+1){\mathcal P}(x).
$$
Let
$$
 Q^{\natural}(x,z)=A_g^{\natural}(z){\mathcal P}^g(x)+\dots+A_0^{\natural}(z).
$$
Let us substitute $Q^{\natural}$ into (\ref{r1}) and use the identities
$$
 {\mathcal P}''(x)=g_2{\mathcal P}(x)+\frac{g_1}{2},\quad  {\mathcal P}'''(x)=g_2{\mathcal P}'(x),
$$
$$
  {\mathcal P}^{(4)}(x)=g_2^2{\mathcal P}(x)+\frac{g_1g_2}{2},\quad  {\mathcal P}^{(5)}(x)=g_2^2{\mathcal P}'(x).
$$
We get
$$
 -\frac{1}{4}{\mathcal P}'(x)(\beta_g{\mathcal P}^g(x)+\dots+\beta_0)=0,
$$
where
$$
 \beta_s=4A_{s+5}g_0^2\frac{(s+5)!}{s!}+4A_{s+4}g_0g_1\frac{(s+4)!}{s!}(2s+5)+
 A_{s+3}\frac{(s+3)!}{s!}(16\alpha_0g_0+g_1(2s+3)(2s+5)
$$
$$
 \left.+8g_0g_2(s(s+4)+5\right)+2A_{s+2}\frac{(s+2)!}{s!}(2s+3)(4\alpha_1g_0+g_1(4\alpha_0+g_2(2s(s+3)+5)))+
$$
$$
 4A_{s+1}(s+1)((s+1)^2(4\alpha_1g_1+g_2(4\alpha_1g_1+g_2(4\alpha_0+g_2(s+1)^2+4z)+
$$
$$
 8A_s(2s+1)\alpha_1g_2(s(s+1)-g(g+1))=0,
$$
we assume that $A_s=0$ at $s<0$ and $s>g$. Hence from $\beta_s=0, 0\leq s<g$ we have
$$
 A_s=\frac{-1}{8(2s+1)\alpha_1g_2(s(s+1)-g(g+1))}\left(4A_{s+5}g_0^2\frac{(s+5)!}{s!}+4A_{s+4}g_0g_1\frac{(s+4)!}{s!}(2s+5)\right.
$$
$$
 +A_{s+3}\frac{(s+3)!}{s!}(16\alpha_0g_0+g_1(2s+3)(2s+5)+8g_0g_2(s(s+4)+5))+
$$
\begin{equation}\label{a1}
 \left.+4A_{s+1}(s+1)((s+1)^2(4\alpha_1g_1+g_2(4\alpha_1g_1+g_2(4\alpha_0+g_2(s+1)^2+4z)\right).
\end{equation}
Let
\begin{equation}\label{a2}
 A_g=\frac{(1\cdot 3\cdot ... \cdot(2g+1)\alpha_1^g(2-g(g+1))\dots((g-1)g-g(g+1))}{4^gg!}.
\end{equation}
Then $Q^{\natural}$ has the form (\ref{Q}) and satisfies (\ref{r1}) and (\ref{eq1}). Theorem 2 is proved.

Let us find the spectral curve.

\vspace{0.4cm}

\noindent{\bf Lemma 1} {\it The spectral curve of $L_4^{\natural}, L_{4g+2}^{\natural}$ is given by the equation
$$
 w^2=\frac{1}{4}(4A_0^2z+A_0(16\alpha_0g_0A_2+48g_0^2A_4+36g_0g_1A_3+3g_1^2A_2+16g_0g_2A_2+A_1((25-8g(g+1))g_0
$$
$$
 +g_1(4\alpha_0+g_2)))-4\alpha_0g_0A_1^2+\frac{1}{4}(4g_0A_2+g_1A_1)^2-2g_0A_1(6g_0A_3+3g_1A_2+g_2A_1)),
$$
where $A_j$ are defined in (\ref{a1}),(\ref{a2}).
}
\vspace{0.4cm}

\noindent {\it Proof.} Let $x_0$ be a zero of ${\mathcal P}(x)$.
Since in the right hand side of (\ref{eq1}) there are only derivatives of order not grater than four, then  we have
$$
 4\tilde{F}_g=\frac{1}{4}(4(z-W)\tilde{Q}^2-4V(\tilde{Q}')^2+(\tilde{Q}'')^2-2\tilde{Q}'\tilde{Q}^{(3)}
 +
 2\tilde{Q}(2V'\tilde{Q}'+4V\tilde{Q}''+\tilde{Q}^{(4)}))|_{x=x_0},
$$
where
$$
 \tilde{Q}^\natural=A^{\natural}_4(z){\mathcal P}^4(x)+\dots+A_0^{\natural}(z).
$$
Using
$$
 ({\mathcal P}'(x))^2=g_0,\quad {\mathcal P}''(x)=\frac{g_1}{2},\quad {\mathcal P}^{(3)}(x)=g_2{\mathcal P}'(x),\quad {\mathcal P}^{(4)}(x)=\frac{g_1g_2}{2}
$$
we get the equation. Lemma 1 is proved.

Let
$$
 H^{\natural}=\partial_x^2+\alpha_1\cos(x)+\alpha_0,\quad  L_4^{\natural}=(H^{\natural})^2-\alpha_1g(g+1)\cos(x)
$$
and we shall use the notation
$$
 <A,B>=AB+BA.
$$

\noindent {\bf Examples:}

\noindent {\bf 1)} $g=1$
$$
 L_6^{\natural}=(H^{\natural})^3+\frac{1}{8}<1-4\alpha_0-12\alpha_1\cos(x),H^{\natural}>+\alpha_1\cos(x),
$$
$$
 F_1^{\natural}(z)=z^3+(\frac{1}{2}-2\alpha_0)z^2+\frac{1}{16}(1-8\alpha_0+16\alpha_0^2-16\alpha_1^2)z+\frac{\alpha_1^2}{4}.
$$
\noindent {\bf 2)} $g=2$, let $\alpha_0=0$
$$
 L_{10}^{\natural}=(H^{\natural})^{5}+\frac{1}{2}<\frac{17}{4}-15\alpha_1\cos(x),(H^{\natural})^3>-
 \frac{15}{2}\alpha_1<\cos(x),(H^{\natural})^2>+
$$
$$
 \frac{1}{2}<1+27\alpha_1^2-60\alpha_1\cos(x)+45\alpha_1^2\cos(2x),H^{\natural}>+\frac{3}{2}\alpha_1(5\cos(x)-12\alpha_1),
$$
$$
 F_2^{\natural}(z)=z^5+\frac{17}{2}z^4+\frac{1}{16}(321-336\alpha_1^2)z^3+
 \frac{1}{4}(34-531\alpha_1^2)z^2+
$$
$$
 (1-189\alpha_1^2+108\alpha_1^4)z+24\alpha_1^2+513\alpha_1^4.
$$

\subsection{Theorem 3}
Let
$$
 Q^{\flat}(x,z)=A^{\flat}_g(z)\wp^g(x)+\dots+A^{\flat}_0(z).
$$
Let us substitute $Q^{\flat}$ into (\ref{r1}). Using the identities
$$
 \wp''(x)=(6\wp^2(x)+g_2\wp(x)+\frac{g_1}{2}),
\quad
 \partial_x^3\wp(x)=(12\wp(x)+g_2)\wp'(x),
$$
$$
 \partial_x^4\wp(x)=(120\wp^3(x)+30g_2\wp^2(x)+(18g_1+g_2^2)\wp(x)+\frac{g_1g_2}{2}+12g_0),
$$
$$
 \partial_x^5\wp(x)=(360\wp^2(x)+60g_2\wp(x)+18g_1+g_2^2)\wp'(x)
$$
we get
$$
 {\mathcal L}_5Q=\wp'(x)(\beta_{g+1}\wp^{g+1}(x)+\dots+\wp(x))=0,
$$
where $\beta_{s+1}=0$ is equivalent to
$$
 16A_s(s+1)((6+8\alpha_1)s+(19+4\alpha_1)s^2+16s^3+4s^4-s_2)+
$$
$$
 8A_{s+1}(2s+3)(4\alpha_0(s+1)(s+2)+g_2(s+1)(s+2)(2s^2+6s+\alpha_1+5)-s_1)+
$$
$$
 4A_{s+2}(4\alpha_0g_2(s+3)^3+g_2^2(s+2)^5+2g_1(s+2)^3(4(s+2)^2+2\alpha_1+5)+4(s+2)z)+
$$
$$
 2A_{s+3}(s+3)(s+2)(2s+5)(4g_0(\alpha_1+2(s^2+5s+9))+g_1(4\alpha_0+g_2(2s^2+10s+13)))+
$$
$$
 A_{s+4}(s+2)(s+3)(s+4)(16\alpha_0g_0+8g_0g_2(s^2+6s+10)+g_1^2(4s^2+24s+35))+
$$
$$
4A_{s+5}g_0g_1(s+2)(s+3)(s+4)(s+5)(2s+7)+
 4A_{s+6}g_0^2\frac{(s+6)!}{(s+1)!}=0.
$$
At $s=g$ and $s=g-1$ this formula has the forms
\begin{equation}\label{u2}
 A_g(g+1)((6+8a_1)g+(19+4a_1)g^2+16g^3+4g^4-s_2)=0,
\end{equation}
$$
 A_g(2g+1)(4a_0g(g+1)+g(g+1)(1+a_1+2g+2g^2)g_2-s_1)+
$$
\begin{equation}\label{u3}
2A_{g-1}g(1-5g^2+4g^4+4a_1(g^2-1)-s_2)=0.
\end{equation}
At
$$
 s_1=\frac{1}{4}g(g+1)(16\alpha_0+5g_2), \quad
 s_2=-4g(g+2)(g^2-1),\quad
 a_1=\frac{1}{4}-2g-2g^2
$$
equations (\ref{u2}),(\ref{u3}) are vanished. From $\beta_{g-1}=0$ we get
$$
 A_{g-2}=\frac{A_{g-1}(4\alpha_0(8g-4)+(16g^3-24g^2+18g-5)g_2)}
 {64(2g^2-3g+1)}+
$$
$$
 \frac{A_{g}(8g^3g_1-g^4g_2^2-g^2(11g_1+4\alpha_0g_2)-4z)}
 {64(2g^2-3g+1)}.
$$
Thus from the equations $\beta_i=0, i>0$ we express $A_{i-1}$ through $A_{g-1}$ and $A_g$. The last equation $\beta_0=0$
takes the form
$$
 A_gP_0(z)+A_{g-1}P_1(z)=0,
$$
where $P_0(z)$ and $P_1(z)$ are some polynomials.
Let
$$
 A_g=-P_1(z),\ A_{g-1}=P_0(z).
$$
So, we find polynomial in $z$ solution of the equation (1).
Theorem 3 is proved.

The equation $w^2=F^{\flat}_g(z)$ of the spectral curve of $L_4^{\natural},L_{4g+2}^{\natural}$ is given in the following lemma.

\vspace{0.4cm}

\noindent{\bf Lemma 2} {\it The spectral curve of $L_4^{\flat}, L_{4g+2}^{\flat}$ is given by the equation
$$
 w^2=\frac{1}{4}(4A_0^2z+A_0(4\alpha_1A_1g_0+48g_0^2+36A_3g_0g_1+3A_2g_1^2+4\alpha_0(4A_2g_0+A_1g_1)+16A_2g_0g_2+
$$
$$
 +A_1g_1g_2)-4a_0A_1^2g_0+\frac{1}{4}(4A_2g_0+A_1g_1)^2-2A_1g_0(6A_3g_0+3A_2g_1+A_1g_2)),
$$
where $A_j$ are defined above.
}
\vspace{0.4cm}

The proof of the Lemma 2 is the same as the proof of the Lemma 1.

Let
$$
 H^{\flat}=\partial_x^2+\alpha_1\wp(x)+\alpha_0.
$$

\noindent {\bf Examples:}

\noindent For the simplification of the formulas we put $g_2=0$.

\noindent {\bf 3)} $g=1$
$$
 L_6^{\flat}=(H^{\flat})^3+\frac{3}{8}<g_1+2\alpha_0\wp(x),H^{\flat}>+
 2\alpha_0g_1+32\alpha_0\wp^2(x),
$$
$$
 F_1^{\flat}=z^3+\frac{3g_1}{2}z^2+(9\alpha_0g_0+4\alpha_0^2g_1+\frac{9}{16}g_1^2)z+4\alpha_0^2g_1^2+
 \frac{27\alpha_0g_0g_1}{4}-16\alpha_0^3g_0,
$$

\noindent {\bf 4)} $g=2, g_1=0$
$$
 L_{10}^{\flat}=(H^{\flat})^5+<30\alpha_0\wp(x)-12\wp^2(x),(H^{\flat})^3>+
 15<16g_0-12\alpha_0\wp^2(x)+16\wp^3(x),(H^{\flat})^2>
$$
$$
 +9<\alpha_0g_0-1850g_0\wp(x)+80\alpha_0^2\wp^2(x)-480\alpha_0\wp^3-11520\wp^2(x),H^{\flat}>-
$$
$$
 36(28\alpha_0^2g_0-1151\alpha_0g_0\wp-6946g_0\wp^2(x)+160\alpha_0^2\wp^3(x)-
 7520\alpha_0\wp^4(x)-30208\wp^5(x)),
$$
$$
 F_2^{\flat}(z)=z^5-387\alpha_0g_0z^3+
 27g_0(16\alpha_0^3+139g_0)z^2+12636\alpha_0^2g_0^2z-243\alpha_0g_0^2(64\alpha_0^3+637g_0).
$$

\section{Appendix I. Spectral curve and eigenfunctions of the Lam\'e operator}

S.P. Novikov proved \cite{N} that if the periodic Schr\"odinger operator
$$
 -\partial_x^2+u(x)
$$
commute with an operator of odd order then it has finite number of gaps in its spectrum (the inverse statement were proved by B.A. Dubrovin \cite{D}).
Such operator is called finite-gap operator. Eigenfunctions of the finite-gap operator are expressed in terms of theta-function of the hyperelliptic
spectral curve via Its formula \cite{Its}.
The famous important finite-gap operator is the Lam\'e operator
$$
 L_2=-\partial_x^2+g(g+1)\wp(x).
$$
It was introduced in 1837. Hermite and Halphen used the following ansatz for eigenfunctions of the Lam\'e operator
$$
 L_2\psi=z\psi,
$$
$$
 \psi=\sum^{g(g+1)/2}_{j=1}a_j(z,\alpha_j)\Phi(x,\alpha_j),
$$
where
$$
 \Phi(x,\alpha)=-\frac{\sigma(x-\alpha)}{\sigma(\alpha)\sigma(x)}e^{\zeta(\alpha)x},
$$
$a_j(z,\alpha_j)$ is some function. With the help of this ansatz Hermite and Halphen studied the cases $g=2,3$.
I.M. Krichever \cite{K2} used some modification of the ansatz to find elliptic solutions of the
Kadomtsev--Petviashvili equation (see also \cite{KE}).
 M.-P. Grosset and A.P. Veselov \cite{GV} proved that
coefficients of the equations for spectral curve of the Lam\'e operator can be found recurrently with the help of elliptic
Bernoulli polynomials.

In the next theorem we find the equation of the spectral curve in the explicit form.
For simplicity of the formulas we assume in this section $g_2=0$, i.e. the Weierstrass function satisfies the equation
$$\
 (\wp'(z))^2=4 \wp^3(z)+g_1\wp(z)+g_0.
$$
Let us introduce the following polynomials in $z$ with the help of recurrent formulas.
Let $A_s=0$ at $s>g$ and $s<0$,
$$
 A_g=\frac{4^g(g^2+g)..(g^2+g-(s+1)s)..(g^2+g-(g-1)g)}{g!},
$$
$$
 A_s=\frac{(s+1)(8A_{s+1}z+A_{s+2}g_1(s+2)(2s+3)+2A_{s+3}g_0(s+2)(s+3))}{4(2s+1)(g^2+g-s(s+1))}.
$$
We also introduce the following function
\begin{equation}\label{Q1}
 Q=A_g\wp^g(x)+\dots+A_0.
\end{equation}

\vspace{0.4cm}
\noindent{\bf Theorem 9} {\it The spectral curve of the Lam\'e operator is given by the equation
$$
 w^2=\frac{1}{4}(4A_0^2z+A_0(4A_2g_0+A_1g_1)-A_1^2g_0).
$$
The eigenfunctions of the Lam\'e operator are
$$
 \psi=\sqrt{Q}e^{iw\int\frac{dx}{Q}}.
$$
}
\vspace{0.4cm}
\noindent{\it Proof.} Our observation is that the solution of the equation (\ref{r4}) for the Lam\'e operator is (\ref{Q1}).
The proof is the same as proof of Theorems 1 and 2. To find eigenfunctions we observe that
the eigenfunctions satisfies the equation
$$
 \psi'-i\chi_0\psi=0,\quad \chi_0=\frac{Q_x}{2iQ}+\frac{w}{Q}.
$$
It follows from
$$
 -\partial_x^2+g(g+1)\wp(x)=(-\partial_x-i\chi_0)(\partial-i\chi_0).
$$
Theorem 9 is proved.

The last theorem gives very effective way to find spectral curve of the Lam\'e operator.

\noindent {\bf Examples:}

\noindent {\bf 5)} $g=1$
$$
 F_1=z^3+\frac{g_1}{4}z-\frac{g_0}{4}.
$$
\noindent {\bf 6)} $g=2$
$$
 F_2=z^5+\frac{21g_1}{4}z^3+\frac{27}{4}g_0z^2+\frac{27}{4}g_1^2z+\frac{81}{4}g_0g_1.
$$

\noindent {\bf 7)} $g=3$
$$
 F_3=z^7+\frac{63}{2}g_1z^5+\frac{297}{2}g_0z^4+\frac{4185}{16}g_1^2z^3+\frac{18225}{8}g_0g_1z^2+
 \frac{3375}{16}(27g_0^2+g_1^3).
$$

\section{Appendix II. Rank two self-adjoint operators and evolution soliton equations}
Let us consider the following system of evolution equations
\begin{equation}\label{ev}
 V_{t}=\frac{1}{4}(6VV_x+6W_x+V_{xxx}),\quad W_{t}=\frac{1}{2}(-3VW_x-W_{xxx}).
\end{equation}

\noindent This system is equivalent to the commutativity condition
$$
 [L_4,\partial_t-{\mathcal A}_3]=0,
$$
where
$$
 L_4=(\partial_x^2+V(x,t))^2+W(x,t),\quad {\mathcal A}_3=\partial_x^3+\frac{3}{2}V(x,t)\partial_x+\frac{3}{4}V_x(x,t).
$$
Following to the papers of I.M. Krichever and S.P. Novikov \cite{KN1,KN2} we call the solution $V(x,t),W(t,x)$ the solution of rank $l$ if for
every $t$ the operator $L_4$ is included in a pair of commuting operators of rank $l$ (hire $l=1$ or $l=2$).
Rank 1 solutions of (\ref{ev}) were found by V.G. Drinfeld and V.V. Sokolov \cite{DS}.
It is a very natural question what is the evolution equation for $Q$ under (\ref{ev})?

\vspace{0.4cm}
\noindent{\bf Theorem 10} (V.N. Davletshina, M.) {\it The polynomial
$$
 Q=(z-\gamma_1(x,t))\dots(z-\gamma_g(x,t))
$$
satisfies the equation
\begin{equation}\label{ev1}
 Q_t=\frac{1}{2}(-3VQ_x-Q_{xxx}).
\end{equation}
}
\vspace{0.4cm}

We shall give the proof of the Theorems 10 and 11 and related statements in an another publication.

The equation (\ref{ev1}) gives a symmetry of the equation (\ref{eq1}).
At $g=1$ the formula (\ref{r2}) gives
$$
 V=\frac{-16F_{1}(\frac{1}{2}(-c_{2}-W))+W^{2}_{xx}-2W_{x}W_{xxx}}{4W^{2}_{x}},
$$
where
$$
 w^2=F_1(z)=z^3+c_2z^2+c_1z+c_0,
$$
is the equation of the spectral curve, and the equation (\ref{ev1}) is reduced to the famous Krichever--Novikov (KN) equation
$$
 W_{t}=\frac{48F_{1}(\frac{1}{2}(-c_{2}-W))-W^{2}_{xx}+2W_{x}W_{xxx}}{8W_{x}}.
$$
KN equation plays an important role in the theory of rank two solutions of the Kadomtsev--Petviashvili (KP) equation at $g=1$. It
would be very interesting to study the equation (\ref{ev1}) in the framework of rank two solutions of KP at $g>1$.

\vspace{0.4cm}
\noindent{\bf Theorem 11} (V.N. Davletshina, M.) {\it
The system (\ref{ev}) has the following rank two solution corresponding to the elliptic spectral curve
$$W(x,t)=-40\wp(bt+x)^{2}-\frac{20}{21}(8b+7g_{2})\wp(bt+x)+c,$$
$$V(x,t)=-10\wp(bt+x)-\frac{2b}{21}-\frac{5g_{2}}{6},$$
where $b,c$ --- const.}
\vspace{0.4cm}

Let us numerically study  the Cauchy problem for (\ref{ev}) with initial data from Corollary 2 and with the rapid decay boundary conditions
$$
 V(x,0)=-\frac{\alpha_1a^2}{\cosh^2(ax)}+\alpha_0,
$$
$$
 W(x,0)=-\frac{s_1a^2}{\cosh^2(ax)}+\frac{s_2a^4}{\cosh^4(ax)},
$$
$$
 V_x(\pm\infty,t)=V_{xx}(\pm\infty,t)=W_x(\pm\infty,t)=W_{xx}(\pm\infty,t)=0.
$$
The behavior of the solution looks like $g$-soliton equation of the Korteveg--de Vris equation (see Fig. 1 --- 12).

It is an interesting problem to find exact solution of the Cauchy problem. It gives soliton deformations of the rank two self-adjoint commuting
differential operators.

\begin{figure}[b]
\centering
\includegraphics[width=0.3\textwidth]{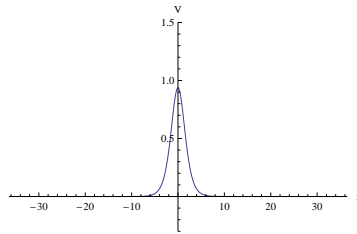}\\
\parbox[]{0.9\textwidth}{\caption{one-soliton solution, $g=1, a=1/2, \alpha_0=0, t=0$}\label{fig1}}
\end{figure}

\begin{figure}
\centering
\includegraphics[width=0.3\textwidth]{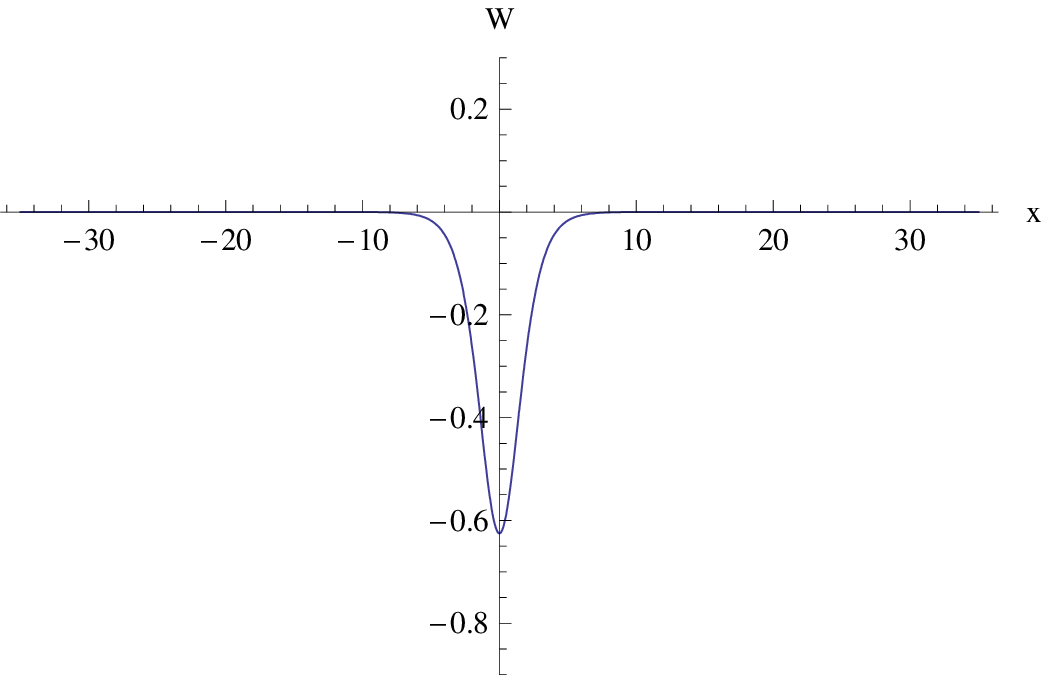}\\
\parbox[]{0.9\textwidth}{\caption{one-soliton solution, $g=1, a=1/2, \alpha_0=0, t=0$}\label{fig1}}
\end{figure}

\begin{figure}
\centering
\includegraphics[width=0.3\textwidth]{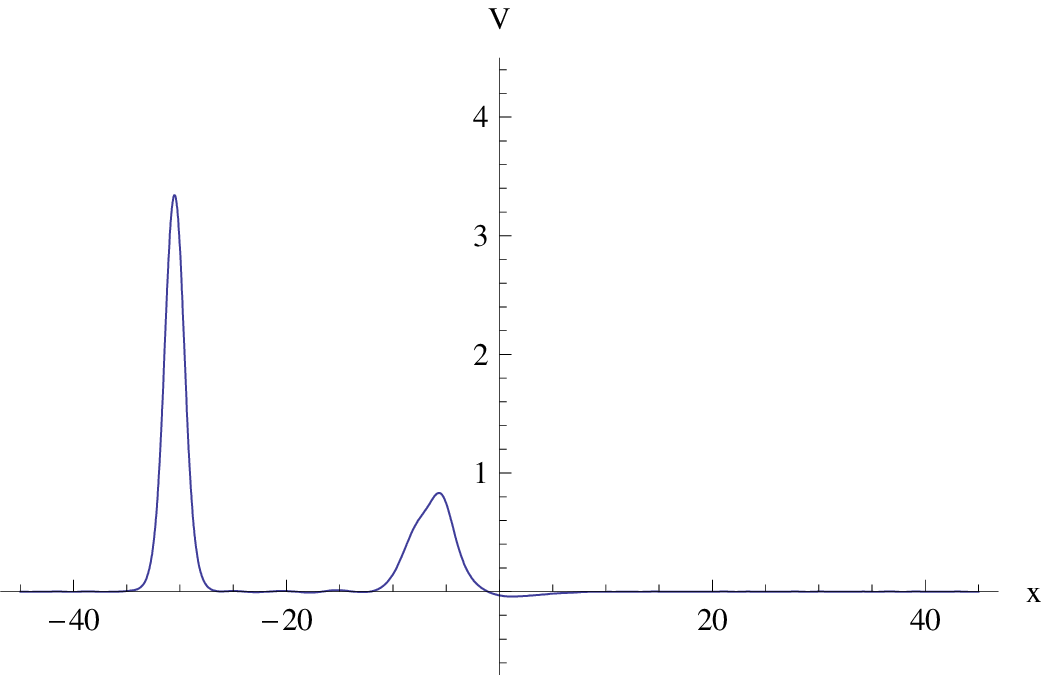}\\
\parbox[]{0.9\textwidth}{\caption{two-soliton solution, $g=2, a=1/2, \alpha_0=0, t=-15$}\label{fig1}}
\end{figure}

\begin{figure}
\centering
\includegraphics[width=0.3\textwidth]{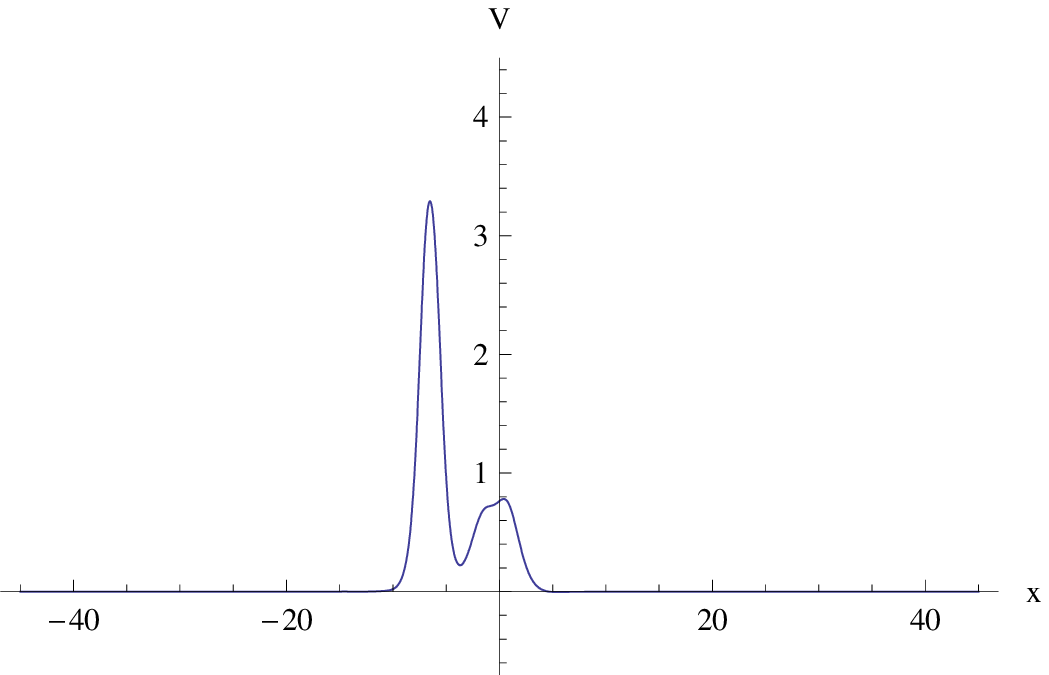}\\
\parbox[]{0.9\textwidth}{\caption{two-soliton solution, $g=2, a=1/2, \alpha_0=0, t=-3$}\label{fig1}}
\end{figure}

\begin{figure}
\centering
\includegraphics[width=0.3\textwidth]{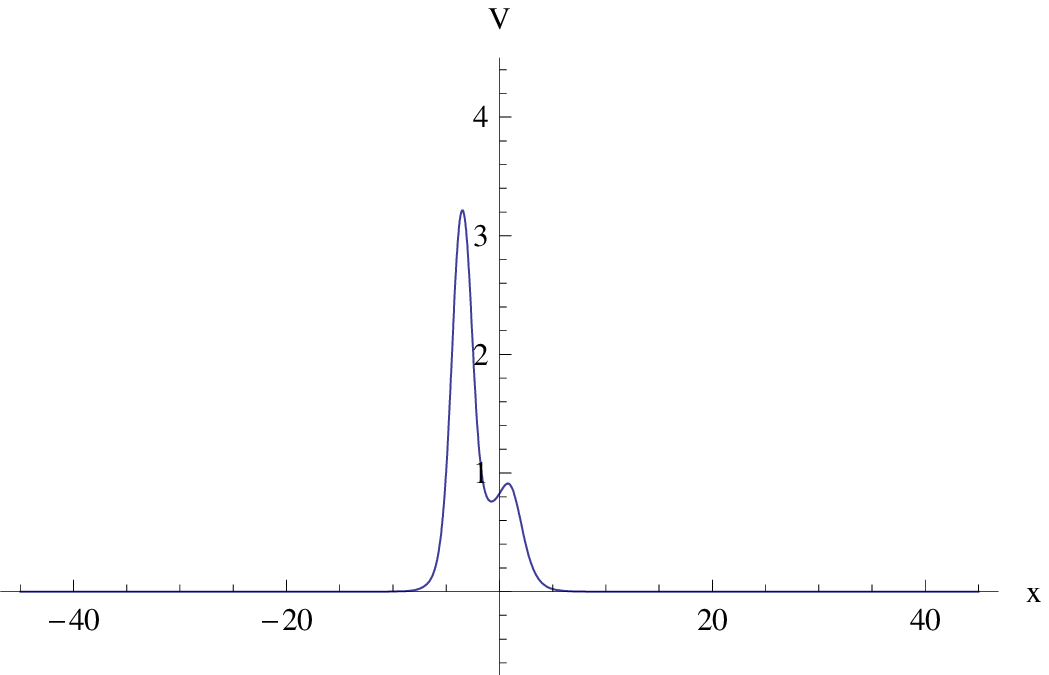}\\
\parbox[]{0.9\textwidth}{\caption{two-soliton solution, $g=2, a=1/2, \alpha_0=0, t=-1.5$}\label{fig1}}
\end{figure}

\begin{figure}
\centering
\includegraphics[width=0.3\textwidth]{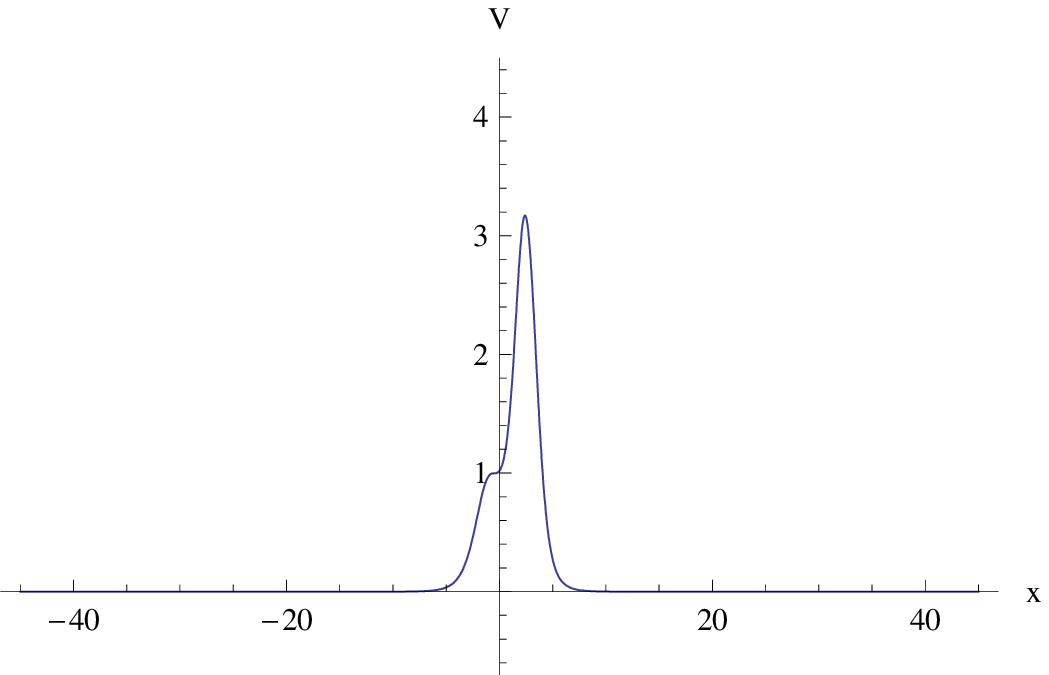}\\
\parbox[]{0.9\textwidth}{\caption{two-soliton solution, $g=2, a=1/2, \alpha_0=0, t=1$}\label{fig1}}
\end{figure}

\begin{figure}
\centering
\includegraphics[width=0.3\textwidth]{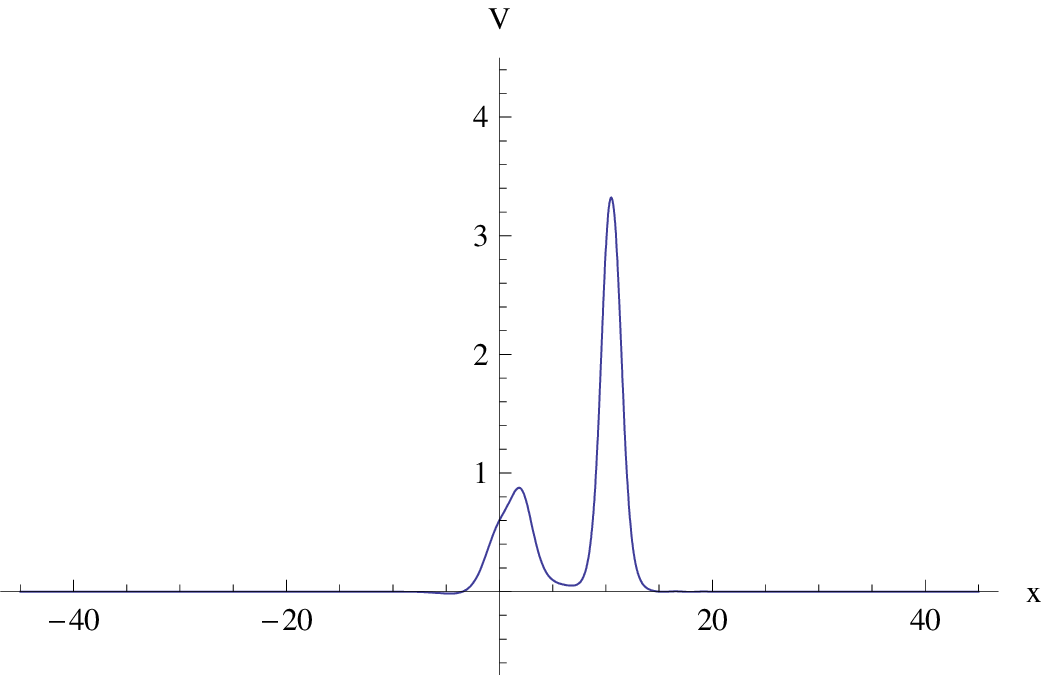}\\
\parbox[]{0.9\textwidth}{\caption{two-soliton solution, $g=2, a=1/2, \alpha_0=0, t=5$}\label{fig1}}
\end{figure}

\begin{figure}
\centering
\includegraphics[width=0.3\textwidth]{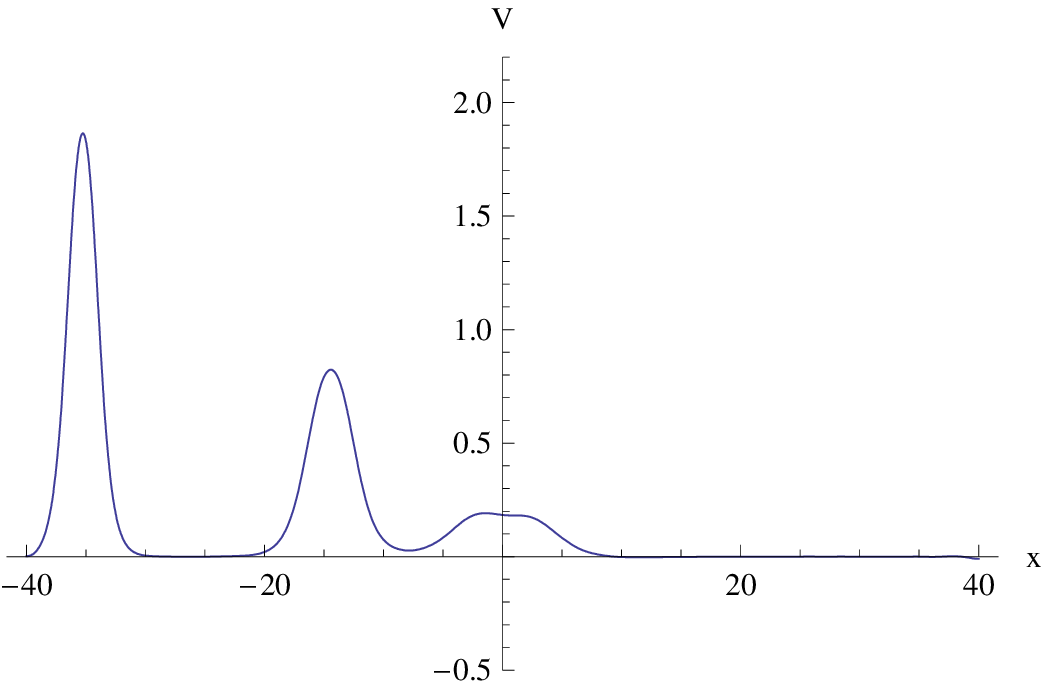}\\
\parbox[]{0.9\textwidth}{\caption{three-soliton solution, $g=3, a=1/4, \alpha_0=0, t=-30$}\label{fig1}}
\end{figure}

\begin{figure}
\centering
\includegraphics[width=0.3\textwidth]{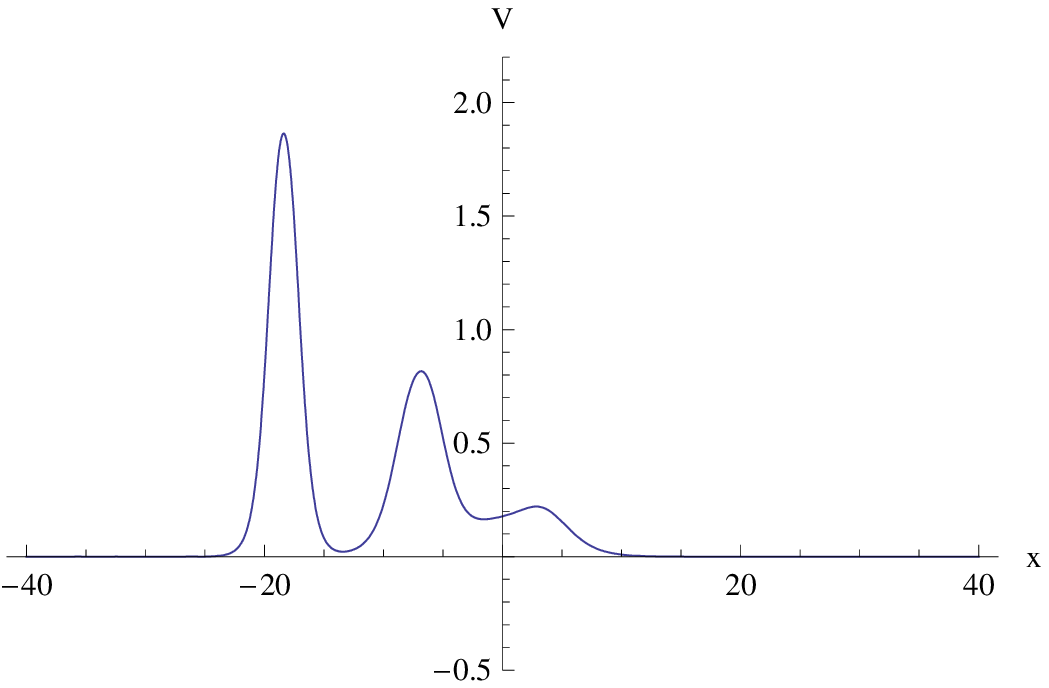}\\
\parbox[]{0.9\textwidth}{\caption{three-soliton solution, $g=3, a=1/4, \alpha_0=0, t=-15$}\label{fig1}}
\end{figure}

\begin{figure}
\centering
\includegraphics[width=0.3\textwidth]{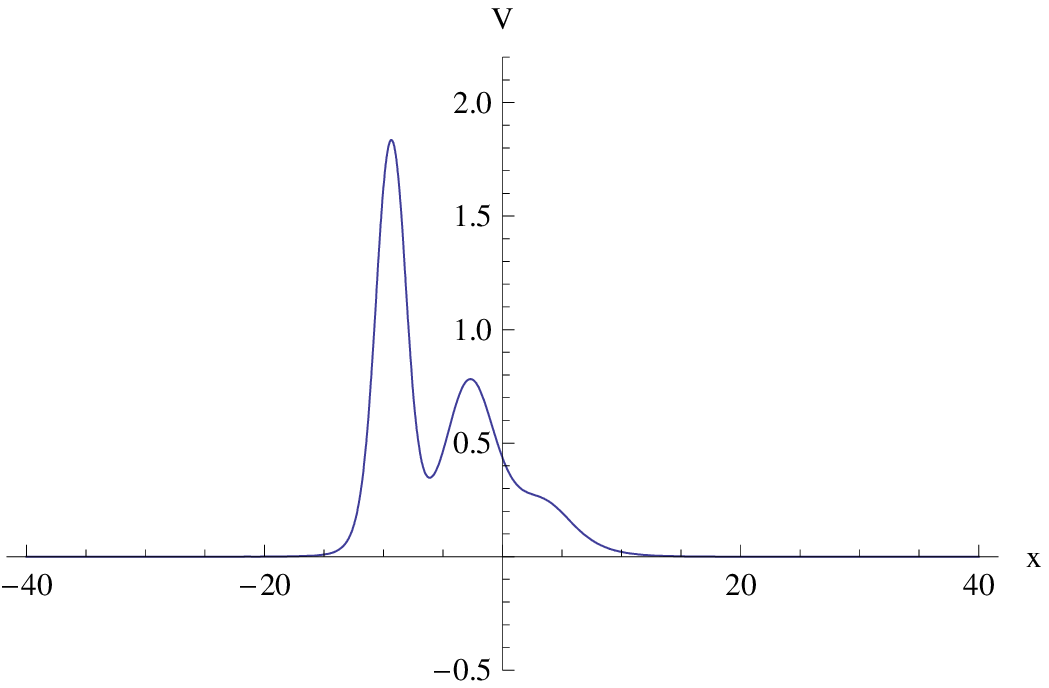}\\
\parbox[]{0.9\textwidth}{\caption{three-soliton solution, $g=3, a=1/4, \alpha_0=0, t=-7$}\label{fig1}}
\end{figure}

\begin{figure}
\centering
\includegraphics[width=0.3\textwidth]{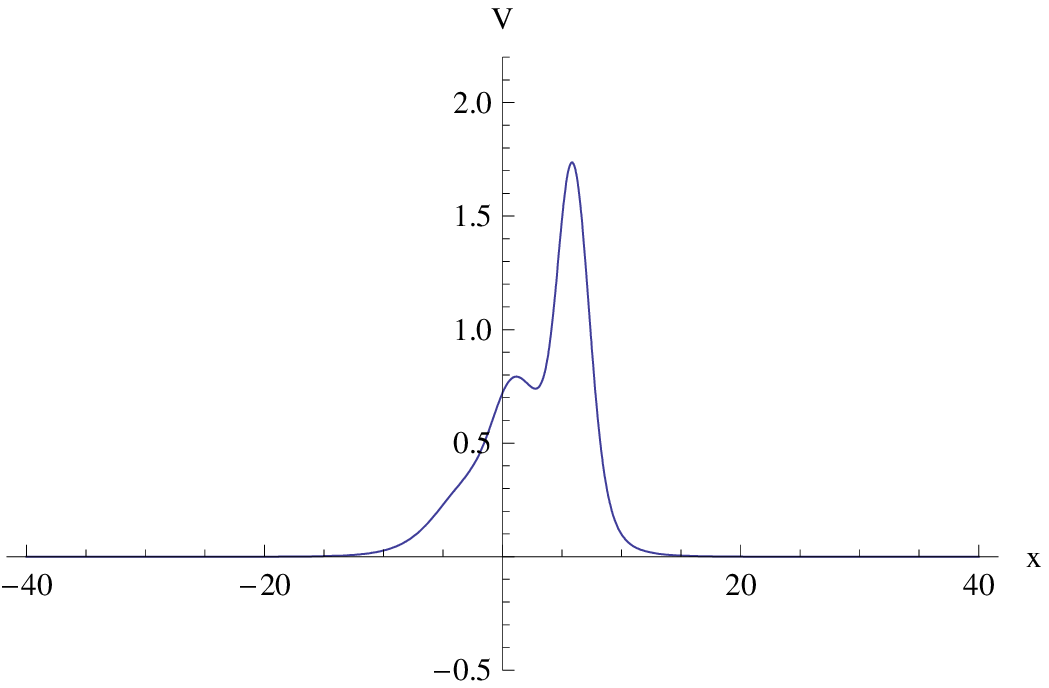}\\
\parbox[]{0.9\textwidth}{\caption{three-soliton solution, $g=3, a=1/4, \alpha_0=0, t=4$}\label{fig1}}
\end{figure}

\begin{figure}
\centering
\includegraphics[width=0.3\textwidth]{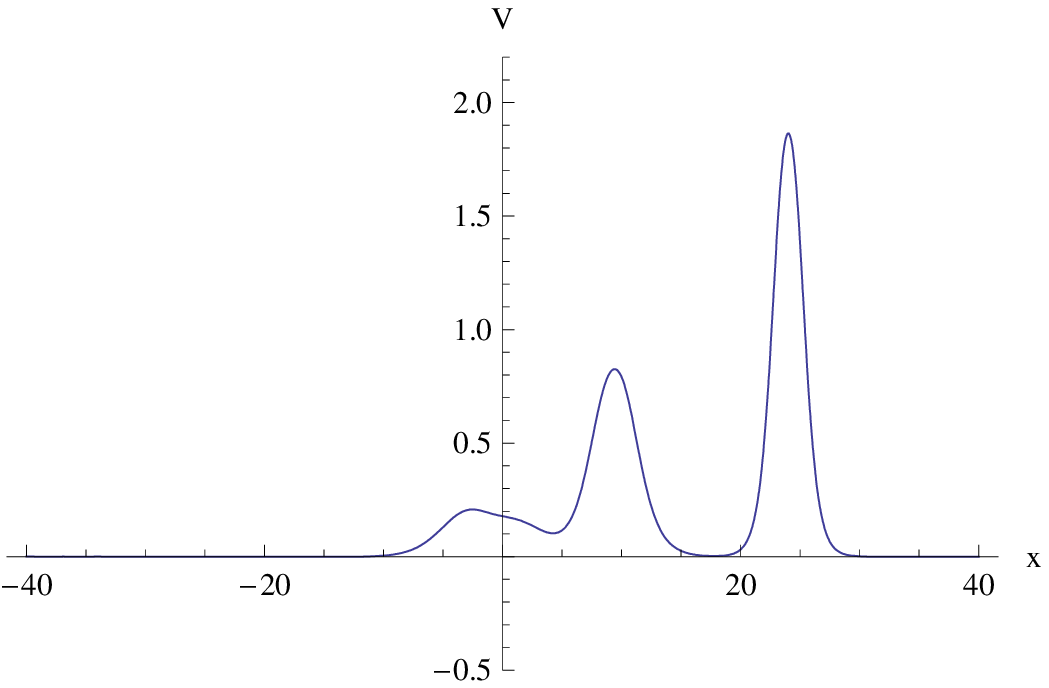}\\
\parbox[]{0.9\textwidth}{\caption{three-soliton solution, $g=3, a=1/4, \alpha_0=0, t=20$}\label{fig1}}
\end{figure}

\vspace{0.4cm}

\noindent Sobolev Institute of Mathematics, Novosibirsk, Russia and

\noindent Laboratory of Geometric Methods in Mathematical Physics, Moscow State University

\noindent e-mail: mironov@math.nsc.ru

\end{document}